\documentclass[aps,prl,twocolumn,showpacs,longbibliography,superscriptaddress]{revtex4-2}
\usepackage[english]{babel}
\usepackage{amsmath}
\usepackage{amsfonts}
\usepackage{amssymb}
\usepackage{graphicx}
\usepackage{amsthm}
\usepackage{bm}
\usepackage[cp1251]{inputenc}

\begin{document}

\title{Long Time Energy Oscillation Between Electron Shell and Nucleus in $^{229}$Th Ions
and Coherent Electron Bridge for Nuclear Quantum Battery}

\author{E.~V.~Tkalya\thanks{https://orcid.org/0000-0002-3521-969X}}

\email{eugene.tkalya@mail.ru}

\affiliation{P.~N.~Lebedev Physical Institute of the Russian
Academy of Sciences, 119991, Leninskiy pr. 53, Moscow, Russia}

\affiliation{National Research Nuclear University MEPhI, 115409, Kashirskoe shosse 31, Moscow, Russia}

\affiliation{Institute of Nuclear and Radiation Physics, Russian Federal Nuclear Center-VNIIEF, 607188,
 Muzrukov Ave 10, Sarov, Nizhny Novgorod region, Russia}

\affiliation{Nuclear Safety Institute of the Russian Academy of Sciences, 115191, Bol'shaya Tulskaya 52, Moscow, Russia}

\date{\today}

\begin{abstract}
The electron shell of the Thorium ion with the $M$1(8.4~eV) transition between levels and the doublet of the $^{229}$Th nucleus ground state with the similar transition represent two qubits spatially inserted one within the other. In the case of relative proximity of the energies of these transitions, weakly damped energy oscillations can be excited between qubits, namely, multiple coherent energy transfer from the electron shell to the nucleus and vice versa. This process in the $^{229}$Th ions does not require resonant (within the width of the levels) coincidence of the transition energies due to the relatively high interaction energy of the electron and nuclear currents. The electron shell ``breathes'', periodically decreasing and increasing in size. The effect can be observed in an ion trap by the intensity of light scattered by thorium-229 ions. This extends the energy range for the $^{229m}$Th$(3/2^+,8.4$~eV) isomer excitation via an electron bridge. Furthermore, the system under consideration is transformed into a nuclear quantum battery when exposed to coherent laser radiation. To ``charge'' the battery, i.e. to excite $^{229m}$Th, one can use developed methods for charging quantum batteries, in particular, coherent excitation of the electron shell followed by coherent transfer of excitation energy to the nucleus (the coherent electron bridge). This opens the way for the design of the $^{229}$Th nuclear quantum battery at the current level of technological development.
\end{abstract}

\maketitle

Since the mid-1970s, it has been known that the $^{229}$Th nucleus has an unusual ground-state doublet \cite{Kroger-76}. One of the doublet's states is the anomalously low-lying nuclear level $3/2^+(E_{\text{is}})$. Its energy $E_{\text{is}}$ is the subject of intensive experimental studies over the last 35 years \cite{Reich-90,Helmer-94,Beck-07,Seiferle-19,Kraemer-23,Tiedau-24,Elwell-24,Yamaguchi-24,Zhang-24-Y,Zhang-24-H}. According to the results of recent measurements, $E_{\text{is}}$ has a value of approximately $8.356$~eV \cite{Tiedau-24,Elwell-24,Zhang-24-Y}. The breakthrough in the accuracy of energy measurements in the works \cite{Kraemer-23, Tiedau-24, Elwell-24, Yamaguchi-24, Zhang-24-Y, Zhang-24-H} is associated with the direct registration of VUV photons emitted by the $^{229}$Th nucleus in the isomeric transition. This was first achieved in the experiment \cite{Kraemer-23}, where isomers of $^{229m}$Th decayed inside a MgF$_2$ crystal. Thus, the experiment \cite{Kraemer-23} confirmed the possibility of observing the $\gamma$ decay of $^{229m}$Th in a large band gap ionic crystals, as predicted earlier in the works \cite{Tkalya-00-JETPL,Tkalya-00-PRC}. The subsequent works \cite{Tiedau-24,Elwell-24,Zhang-24-Y,Zhang-24-H} followed the same path by implanting $^{229m}$Th nuclei inside crystals, where the energy and the half-life of the transition were measured in the process of $^{229}$Th resonant excitation of by laser radiation.

The increased interest in the low-lying isomeric state of $^{229m}$Th~($3/2^+$, $E_{\text{is}}$) and the isomeric transition $3/2^+(E_{\text{is}}) \rightarrow 5/2^+(0.0)$ is explained mainly by the possibility of creating a nuclear frequency (time) standard \cite{Tkalya-96,Peik-03,Rellergert-10,Campbell-12,Peik-15,Beeks-21} and a nuclear laser \cite{Tkalya-11,Tkalya-13}, as well as the search for new physics beyond the standard model \cite{Flambaum-06,Litvinova-09,Berengut-09,Safronova-18-RMP,Peik-21} and a number of fundamental physical phenomena (see in \cite{Dykhne-96,Pachucki-01,Tkalya-11,Tkalya-16-PRA,Tkalya-18-PRL} and references therein).

In atomic-nuclear physics, cases are known where the multipolarities of some nuclear and atomic transitions are identical, and their energies coincide within the radiative linewidths. This occurs, for example, in $^{197}$Au and $^{193}$Ir, where the energies of the nuclear $M$1 transitions, 77.351~keV and 73.04~keV, respectively, are close to the correspondent $X$-ray $M$1~($M_1 \rightarrow{}K$) transitions in the atomic shells of Au and Ir. In $^{197}$Au and $^{193}$Ir, the interaction energy of the electron and nuclear currents is several orders of magnitude smaller than the widths of the atomic levels involved in the process \cite{Tkalya-07}. Therefore, energy transfer from an atomic shell to a nucleus (the well-known NEET process from Nuclear Excitation by Electron Transition \cite{Morita-73}) or from a nucleus to an atomic shell (the inverse NEET \cite{Tkalya-94-NEET-JETPLett}) occurs once with a very low probability.

In this paper, we consider another situation that can arise in $^{229}$Th ions when the $M1$ transitions between the electronic states are close to the energy of the low-lying isomeric state in the $^{229}$Th nucleus. In general, this is a model situation. Nevertheless, according to data from \cite{Dzuba-25-PRA-L,Dzuba-25-PRA,Yu-25,Koziol-25}, this is entirely possible in $^{229}$Th$^{+,2+,6+,39+}$. Other thorium ions cannot be ruled out either: NEET is a component of the electron bridge, a process of efficient excitation and decay of the $^{229m}$Th isomer which requires the same conditions and was studied for many Thorium ions \cite{Tkalya-92-JETPL, Tkalya-92-SJNP, Kalman-94, Porsev-10-PRL, Karpeshin-17, Bilous-18-NJP, Borisyuk-19-PRC, Bilous-20, Dzyublik-20, Dzuba-25-PRA-L,Dzuba-25-PRA,Shigekawa-26,Wense-26}.

We use a simplified scheme of two qubits -- an atomic qubit with a transition energy between isolated levels $\omega_A$ and a nuclear qubit with a transition energy between the ground and isomeric states $\omega_N = E_{\text{is}}$ (the system of units is $\hbar = c = 1$). A nuclear qubit is ``embedded'' within an atomic qubit. The qubits are in the multipole-static inner region of each other and the electromagnetic field of the multipoles does not experience any significant difference when interacting with the nucleus or the atomic shell, despite the large difference in their sizes.

At the first step, we assume that the energies $\omega_A$ and $\omega_N$ coincide within the electron transition width. This provides an upper bound on the characteristics of the process under consideration. In the next step we will show that the working range of the process significantly exceeds the width of the levels.

The process of energy transfer we will describe by the Limbland equation for the density matrix for two qubits $\rho$ \cite{Ficek-02,Gonzalez-Tudela-11,Ren-17}
\begin{eqnarray}
\partial_t\rho &=& i[\rho,H] + \sum_{i,j=A,N \atop{} i\neq{}j}
\gamma_{ij}\left(S^-_i\rho{}S^+_j - \frac{1}{2}\left\{S^+_i S^-_j,\rho\right\} \right) + \nonumber \\
&&\sum_{i=A,N} \Gamma_i \left(S^-_i\rho{}S^+_i - \frac{1}{2}\left\{S^+_i S^-_i,\rho\right\} \right)
,
\label{eq:Lindblad}
\end{eqnarray}
where the curly braces denote the anticommutator, $S^-_i = |g_i\rangle\langle{}e_i|$ and $S^+_i = |e_i\rangle\langle{}g_i|$ are jump operators for atomic (if $i=A$) and for nucleus (if $i=N$) qubits. In the set $\{|e_A\rangle|e_N\rangle$, $|e_A\rangle|g_N\rangle$, $|g_A\rangle|e_N\rangle$, $|g_A\rangle|g_N\rangle\}$ jump operators have the form
\begin{equation*}
S^+_A={S_A^-}^\dag = \left(
\begin{array}{cccc}
0&0&1&0\\
0&0&0&1\\
0&0&0&0\\
0&0&0&0\\
\end{array}
\right),\,
S^+_N={S_N^-}^\dagger = \left(
\begin{array}{cccc}
0&1&0&0\\
0&0&0&0\\
0&0&0&1\\
0&0&0&0\\
\end{array}
\right).
\end{equation*}
The functions $|g_i\rangle$ and $|e_i\rangle$ are the ground and excited states of the qubits, $\Gamma_i$ in Eq.~(\ref{eq:Lindblad}) is the total (in our case, radiative) width of the states $|e_i\rangle$, and the parameter $\gamma_{ij}$ describes the influence of the spontaneous decay of the nuclear (atomic) state on the spontaneous decay of the atomic (nuclear) excited state. In general, the terms with the parameters $\Gamma_i$ and $\gamma_{ij}$ in Eq.~(\ref{eq:Lindblad}) describe the incoherent part of the processes.

The Hamiltonian $H$ in Eq.~(\ref{eq:Lindblad}) describes the process of coherent interaction between atomic and nuclear qubits. The Hamiltonian has the standard form \cite{Ficek-02,Gonzalez-Tudela-11,Ren-17}
\begin{equation}
H = \sum_{i=A,N} E_i S^+_iS^-_i + \sum_{i,j=A,N \atop{} i\neq{}j}
g_{ij} S_i^+ S^-_j
,
\label{eq:Hamiltonian}
\end{equation}
where $E_i$ is the energy of qubits in the excited state, that is, $E_A=\omega_A$, $E_N\equiv{}E_{\text{is}}=\omega_N$ (the energies of the ground state of the electron shell and the nucleus are taken to be zero), $g_{ij}$ is a term describing the coherent interaction of the nuclear and atomic magnetic dipoles of the transition.

\begin{figure}
 \includegraphics[angle=0,width=0.98\hsize,keepaspectratio]{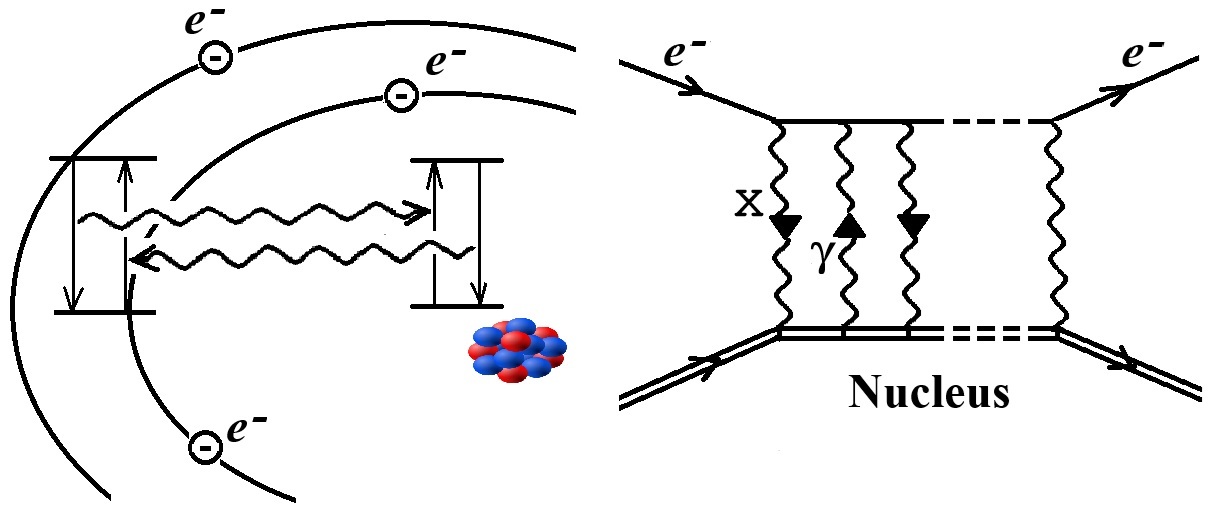}
 \caption{Scheme of energy exchange between the electron shell and the nucleus from a quantum-electrodynamical point of view.}
 \label{fig:Fig1}
\end{figure}

The NEET process of near resonant energy transfer from the electron shell to the nucleus and vice versa in Fig.~\ref{fig:Fig1} is an analogue of the well-known F\"{o}rster resonance energy transfer \cite{Novotny-06}. A detailed theory of NEET within the framework of perturbation theory for quantum electrodynamics was developed in \cite{Tkalya-92-NPA,Tkalya-92-JETP,Tkalya-94,Tkalya-07,Dzyublik-11,Dzyublik-13}. The one-time energy transfer process corresponds to the second-order diagram with a virtual photon (a part of the Feynman diagram in Fig.~\ref{fig:Fig1}) and a Hamiltonian of the interaction of the electron $j^{\mu}_{fi}({\bf{r}})$ and nuclear $J^{\nu}_{fi}({\bf{R}})$ currents
\begin{equation*}
H_{\text{int}}=\int{}d^3rd^3Rj^{\mu}_{fi}({\bf{r}})
D_{\mu\nu}(\omega_N,{\bf{r}}-{\bf{R}})J^{\nu}_{fi}({\bf{R}}) \,,
\label{eq:Hint}
\end{equation*}
where
$D_{\mu\nu}(\omega_N,{\bf{r}}-{\bf{R}}) = g_{\mu\nu}e^{i\omega_N|{\bf{r}}-{\bf{R}}|}/|{\bf{r}}-{\bf{R}}|
$
is the photon propagator, $j^{\mu}_{fi}({\bf{r}}) = -e\bar{\psi}_f({\bf{r}})\gamma^{\mu}{\psi}_i({\bf{r}})$,
$J^{\nu}_{fi}({\bf{R}}) = e\Psi^+_{f}({\bf{R}})\hat{J}^{\nu}\Psi_i({\bf{R}})$,
$\gamma^{\mu}$ are the Dirac matrices, $\hat{J}^{\nu}$ is the operator of the nuclear electromagnetic transition, $\psi_{i,f}$ and $\Psi_{i,f}$ are, respectively, the wave functions of the electron and the nucleus in the initial $i$ and final $f$ states.

%
%

The Green's function of a photon is expanded in multipoles. In the case of the $M1$ transition the interaction energy of an unpolarized nucleus and an atomic shell is \cite{Tkalya-92-NPA,Tkalya-92-JETP,Tkalya-07}
\begin{equation}
E_{\text{int}} \simeq {i}e^2 \frac{\omega_N^{2}}{M_N} {\cal {R}}_1^{M}(\omega_N)
\sqrt{B_{\text{W.u.}}(M1;i\rightarrow{}f)} \,,
\label{eq:EintM1}
\end{equation}
where ${\cal{R}}_1^{M}$ is the electron radial matrix element
\begin{eqnarray}
{\cal{R}}_1^{M}(\omega_A) &=& (\kappa_i+\kappa_f) \int_0^{\infty}dr\,r^2 h_1^{(1)}(\omega_Ar) \times \nonumber\\ &&[g_i(r)f_f(r)+f_i(r)g_f(r)],
\label{eq:R_M1}
\end{eqnarray}
and now $g_{AN}$ and $g_{NA}$ are $\Re(E_{\text{int}})$, $\gamma_{AN}$ and $\gamma_{NA}$ are $\Im(E_{\text{int}})$.

In Eqs.~(\ref{eq:EintM1})--(\ref{eq:R_M1}) $\kappa=(l-j)(2j+1)$, $g(r)$ and $f(r)$ are respectively the large and small components of the electrons wave function, $j_1(x)$ and $h^{(1)}_1(x)$ are the spherical Bessel and Hankel functions, respectively, $M_N$ is the nucleon mass, $B_{\text{W.u.}}(M1;i\rightarrow{}f)$ is the nuclear reduced probability in Weisskopf unit of $B(M1;W) = (45/8\pi)\mu_N^2$, $\mu_N$ is the nuclear magneton. The experimental value of $B_{\text{W.u.}}(M1;5/2^+\rightarrow 3/2^+)$ was obtained in \cite{Tiedau-24, Zhang-24-Y} and was 0.022. The corresponding width of the radiative transition in the bare nucleus is $\Gamma_N = 2.7\times10^{-19}$~eV, and the characteristic magnetic moment is $\mu_{\text{is}\rightarrow\text{gr}} \simeq 0.2\mu_N$.

The diagram in Fig.~\ref{fig:Fig1} may have a large number of vertices. In the near resonance case, the characteristic energy denominators in the electron and nuclear propagators have the order of $\Gamma_A$ and $\Gamma_N$. In the matrix elements in the numerator, we are dealing with bound electron states localized in a region of the order of the Bohr radius $r\lesssim{}a_B$. For the considered transition energy one has $\omega_{A(N)}a_B\ll 1$. Thus, in the region of integration of the electron matrix elements ${\cal{R}}_1^{M}$ in Eq.~(\ref{eq:R_M1}), the function $h^{(1)}_1(\omega_A{}r)$ reaches large values and compensates for the smallness introduced by the vertices in the diagram in Fig.~\ref{fig:Fig1}. As a result the process amplitude is not small. The single-level approximation for the ion and nucleus allows us to cut the diagram with vertical lines and represent the process amplitude as the product of approximately equal amplitudes for the forward and inverse NEET processes, each with a probability close to 1. This creates the possibility of multiple energy oscillations in the system.

To numerically estimate the interaction energy $E_{\text{int}}$, we consider two cases: a highly charged ion Th$^{39+}$ and a low-charge ion Th$^{+}$. In the paper \cite{Koziol-25} the width of the atomic $M1(E_{\text{is}})$ transition $\Gamma_A = 8.7\times10^{-12}$~eV and the probability of the NEET process $W_{NEET} = 2.5\times10^{16}$~s$^{-1}$ were calculated for Th$^{39+}$. The characteristic magnetic moment of the transition, reconstructed from the value of $\Gamma_A$, amounted to a significant value of $|{\boldsymbol{\mu}}_A| \simeq \mu_B$. The interaction energy follows from $W_{NEET}$: $E_{\text{int}}=(W_{NEET}\Gamma_A/4)^{1/2} \simeq 6\times10^{-6}$~eV. (The interaction energy for low-energy transitions depends almost exclusively on the imaginary part of the atomic matrix element $\Im[{\cal{R}}_1^{M}(\omega_A)]$ in Eq.~(\ref{eq:R_M1}) \cite{Tkalya-92-NPA}, which immediately gives the presented value of $E_{\text{int}}$.)

For the low-charge ion Th$^{+}$, we consider the electron $M1$ transition between the states $7S_{1/2}$ and $8S_{1/2}$ which energy is close to the energy of the nuclear transition. (Transitions between $S_{1/2}$ states are the most intense among $M1$ transitions.) Calculation within the relativistic Hartree-Fock-Slater approximation yields for the imaginary part of the matrix element $\Im[{\cal{R}}_1^{M}(\omega_N)] \simeq -5\times 10^{5}$. This value of the matrix element can be considered reliable:  similar calculations reproduces the experimental data on the probability of internal electron conversion for the isomer $^{229m}$Th \cite{Strizhov-91,Seiferle-17}. Thus, the coherent part of the interaction energy $g_{AN} = \Re[E_{\text{int}}] \simeq 10^{-6}$~eV, that is, the same order of magnitude as in the Th$^{39+}$ ion.

The incoherent part of the interaction energy $\gamma_{AN,NA} = \Im(E_{\text{int}})$ is relatively small in both cases because of $\Re[{\cal{R}}_1^{M}(\omega_A)]\ll\Im[{\cal{R}}_1^{M}(\omega_A)]$. For the Th$^{39+}$ ion the value of $\gamma_{AN,NA}\simeq 10^{-15}$~eV can be easily restored by taking $\Re[{\cal{R}}_1^{M}] \simeq -1.2\times10^{-5}$ from the known width of the radiation $M1$ transition. For the Th$^+$ ion according to \cite{Gossel-13}, the electron core polarization for $M$1 amplitudes gives a contribution several orders of magnitude larger than that given by relativistic Hartree-Fock calculations alone. The reduced matrix elements for the operator of interaction of the magnetic moment of the atomic transition ${\boldsymbol{\mu}}_A$ with the magnetic field of the emitted photon {\boldmath${B}$}$_X$ is \cite{Savukov-99,Gossel-13}
\begin{eqnarray*}
\left\langle{}f\| {\boldsymbol{\mu}}_A\cdot{\boldsymbol{B}}_X \|{i}\right\rangle &=&
(\kappa_i +\kappa_f)\left\langle -\kappa_f\|{C_1}\|\kappa_i\right\rangle  \\
&&\times \int_0^{\infty}\frac{3}{\omega_A} j_1(\omega_A r)[g_if_f+g_ff_i]r^2dr\,,
\label{eq:muB}
\end{eqnarray*}
where $C_1$ is the normalized spherical harmonic \cite{Varshalovich-88}. The calculation carried out in \cite{Gossel-13} gave the following value $\left\langle{}f\| {\boldsymbol{\mu}}_A\cdot{\boldsymbol{B}}_X \|{i}\right\rangle = -2549\times10^{-5}\mu_B$. From here, for the real part of the radial matrix element in Eq.(\ref{eq:R_M1}), we obtain $\Re[{\cal{R}}_1^{M}(\omega_A)] \simeq -1.0\times 10^{-6}$ and, as a consequence, a very small value of the incoherent part of the interaction energy $\gamma_{AN} = \Im{}E_{\text{int}} \simeq 2\times 10^{-18}$~eV.

The probability of $M1$ emission in the $8s_{1/2} \rightarrow 7s_{1/2}$ transition is calculated from the formula
\begin{equation*}
\Gamma^{M1}_{\text{rad}} = (\omega_A^3/6)|\left\langle{}f\| {\boldsymbol{\mu}}_A\cdot{\boldsymbol{B}}_X \|{i}\right\rangle|^2
\simeq 4.4\times10^{-16}\,\,{\text{eV}}.
\end{equation*}
Hence, the characteristic magnetic moment of the transition is $|{\boldsymbol{\mu}}_A| \simeq 10^{-2}\mu_B$, which is indeed significantly greater than the nuclear magnetic moment of the transition between the ground and isomeric states.

So, $\gamma_{AN,NA} \simeq 10^{-15}$~eV -- $10^{-18}$~eV in a wide range of ionization of Thorium atoms. Thus, the broadening of the widths of the atomic states due to the incoherent part of the interaction is negligible. (For comparable values of $\Gamma_A$ and $\Gamma_N$, the incoherent part of the interaction energy can slow down the decay rate of the system at large times: the system attenuates according to the law
$$\frac{1}{2}e^{-\left(\frac{\Gamma_A+\Gamma_N}{2}-\gamma_{AN,NA}\right)t}$$
because decay occurs via the antisymmetric state \cite{Ficek-02,Gonzalez-Tudela-11}
$$\frac{1}{\sqrt{2}}\left(|A_eN_g\rangle -|A_gN_e\rangle\right).$$
However, for the Thorium ions under consideration this is not relevant due to the smallness of $\Gamma_N$ and $\gamma_{AN,NA}$ compared to $\Gamma_A$.

For order-of-magnitude estimates in further calculations, we neglect the nuclear width of the transition $\tilde{\Gamma}_N$ compared to the atomic $\Gamma_A = \Gamma^{M1}_{\text{rad}}$, and the insignificant difference between $g_{AN}$ and $g_{NA}$, i.e., we set $g_{AN}\simeq g_{NA} = g$. In this case, the system Eq.(\ref{eq:Lindblad}) is solved analytically. In particular, if at the initial moment the electron shell is in the excited state $A_e$, and the nucleus is in the ground state $N_g$, then the probability to find the nucleus in the excited state $N_e$ and the ion in the ground state $A_g$ when the typical for the NEET process conditions $g \ll \Delta{}E \lesssim \Gamma_A$ ($\Delta{}E$ is energy detuning between the energies of the atomic and nuclear transitions) are met is described by the relation
\begin{eqnarray}
\rho_{A_gN_e}(t) &=& \frac{g^2}{\Delta{}E^2 + \Gamma_A^2/4} \biggl( \left(1-e^{-\Gamma_A{}t/2} \right)^2  \nonumber\\
&& +\, 4e^{-\Gamma_A{}t/2}  \sin^2\bigl(\Delta{}E{}t/2\bigl) \biggl),
\label{eq:P_NEET}
\end{eqnarray}
the main term of which is well known from the NEET theory \cite{Tkalya-92-NPA}. (For brevity, this paper uses abbreviated notations for the four elements of the density matrix: $\rho_{A_gN_e}=\rho_{A_{gg}N_{ee}}$, $\rho_{A_eN_g}=\rho_{A_{ee}N_{gg}}$, $\rho_{A_gN_g}=\rho_{A_{gg}N_{gg}}$, and $\rho_{A_eN_e}=\rho_{A_{ee}N_{ee}}$.)

\begin{figure}
 \includegraphics[angle=0,width=0.9\hsize,keepaspectratio]{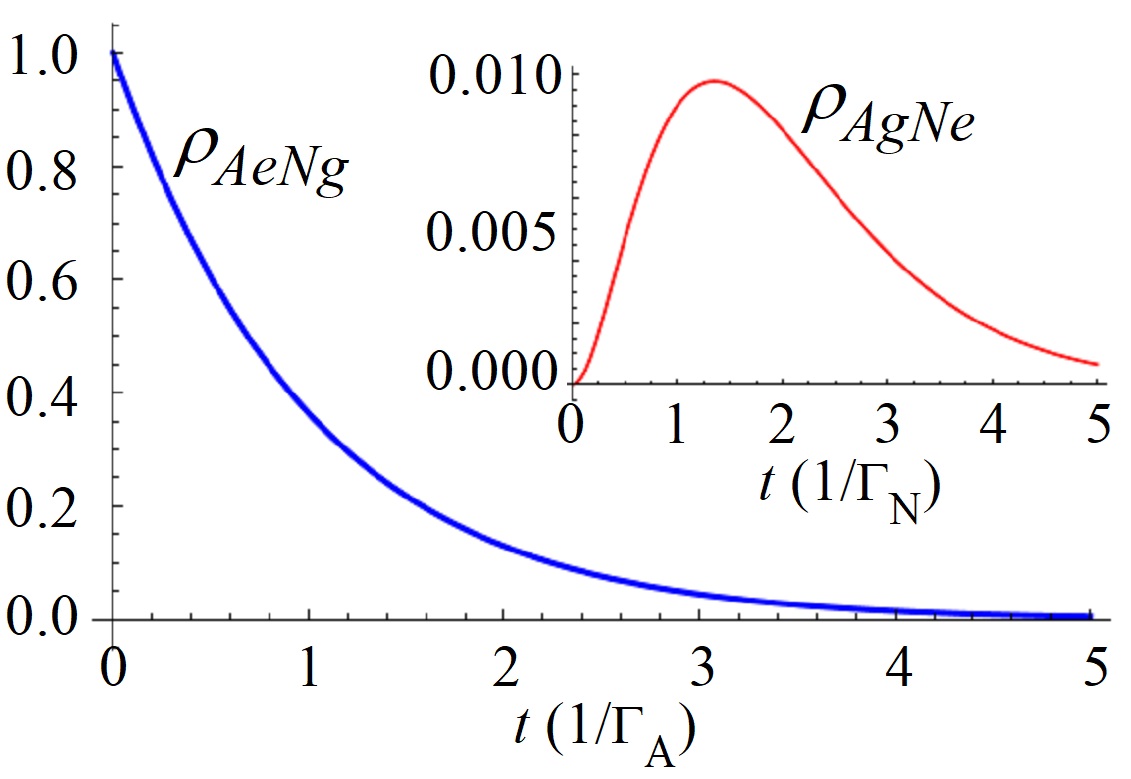}
 \caption{One-time NEET process for $\Gamma_A = 10g$ and $\Gamma_N = 5g$.}
 \label{fig:Fig2}
\end{figure}

In the described by Eq.~(\ref{eq:P_NEET}) atomic-nuclear system (for example, $^{197}$Au and $^{193}$Ir) a single transfer of energy from the electron shell to the nucleus occurs with a low probability proportional to $g^2/(\Delta{}E^2 + \Gamma_A^2/4)$ \cite{Tkalya-92-NPA}. A typical NEET process for such a case is shown in Fig.~\ref{fig:Fig2}). In the case of a large mismatch $\Delta{}E$ (see, for example, NEET in $^{189}$Os \cite{Tkalya-92-NPA}), the probability of the NEET process rapidly decreases as $g^2/\Delta{}E^2$.

%
%

\begin{figure}
 \includegraphics[angle=0,width=0.9\hsize,keepaspectratio]{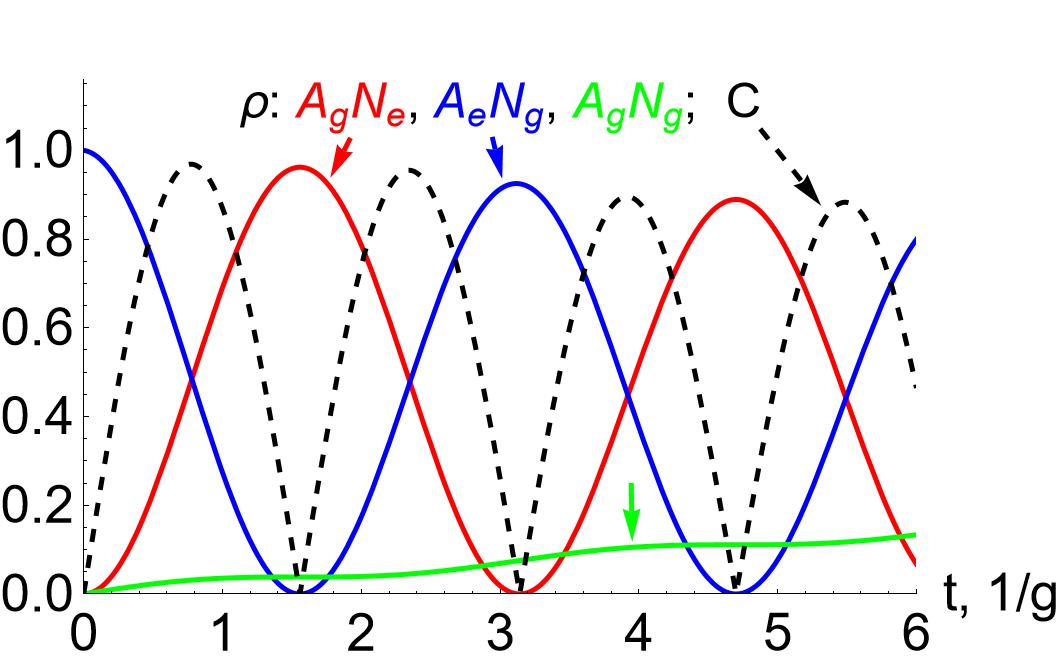}
 \caption{The probability to find the system in the states $A_eN_g$ --- blue, $A_gN_e$ --- red, $A_gN_g$ --- green for the ratio of constants $\Gamma_A = g/20$ and under the initial condition $A_eN_g(0)=1$. $C(t)$ is concurrence.}
 \label{fig:Fig3}
\end{figure}

In the $^{229}$Th ions, the interaction energy $g = \Re(E_{\text{int}})$ significantly exceeds the width $\Gamma_A$. Therefore, at resonance, when the condition $\Delta{}E \ll \Gamma_A \ll{} g$ is satisfied, the probability of excitation of the nucleus as a function of time has the form
\begin{equation*}
\rho_{A_gN_e}(t) = e^{-\Gamma_A t/2} \frac{\sin^2\bigl(gt\sqrt{1-(\Gamma_A/4g)^2}\bigl)}{1-\bigl(\Gamma_A/4g\bigl)^2}.
\label{eq:PNeAgRes}
\end{equation*}
In Fig.\ref{fig:Fig3} this is a red curve oscillating with a frequency of $\nu \simeq 10^9$~s$^{-1}$, which is typical for $^{229}$Th$^{39+}$. Also graphs are presented for the probability to find the system in the initial state and in the state when the electron shell and nucleus are not excited. The oscillations continue for a time $t\simeq 1/\Gamma_A \simeq 10^{-5}-10^{-4}$~s. This is reminiscent of an atom in a resonator. In this case, the electron shell of the ion plays the role of a kind of resonator for the nucleus, which emits a virtual photon. Note that, according to \cite{Koziol-25}, here, the electron shell has no relaxation channel other than the radiative one mentioned above.

%
%

Entanglement is spontaneously formed in the system under consideration in accordance with the expression for concurrence \cite{Gonzalez-Tudela-11}
\begin{eqnarray*}
C(t)&=&\bigl((\rho_{A_{ge}N_{eg}}+\rho_{A_{eg}N_{ge}})^2+[\Im(\rho_{A_{ee}N_{gg}}+ \nonumber\\
&& \rho_{A_{ge}N_{eg}}-\rho_{A_{eg}N_{ge}}-\rho_{A_{gg}N_{ee}})]^2\bigl)^{1/2}.
\end{eqnarray*}
As expected, with our choice of parameters $\gamma_{AN}= \gamma_{NA}=0$ the concurrence in Fig.\ref{fig:Fig3} oscillates and reaches maxima at the points $\rho_{A_{g}N_{e}}=\rho_{A_{e}N_{g}}$.

Another feature of the energy exchange process in the $^{229}$Th ions is its behavior at large energy detunings, that is, at $\Gamma_A \ll \Delta{}E \ll g$. This case is described by the equation
\begin{equation*}
\rho_{A_gN_e}(t) = \left(1-\frac{\Delta{}E^2}{4g^2}\right) e^{-\Gamma_A t/2} \sin^2\left(gt\left(1+\frac{\Delta{}E^2}{8g^2}\right)\right).
\label{eq:PNeAgNRES}
\end{equation*}
In contrast to the standart NEET process discussed above, here the probability of nuclear excitation practically does not decrease with increasing $\Delta{}E$, even when the detuning is orders of magnitude greater than the width $\Gamma_A$. (In $^{229}$Th$^{39+}$  detuning between transitions can down to the level of $\Delta{}E\simeq 10^{-6}$~eV or $\Delta{}E \simeq 10^5\Gamma_A$.) The system still oscillates rapidly with a frequency of $\nu = g/2\pi$, and the probability of excitation of the nucleus is close to 1. That is, there is no need a resonant coincidence of the energies of the atomic and nuclear transitions for the existence of oscillations at large $g$. This does not contradict the law of the energy conservation, since the time for which the nucleus is excited $\Delta{}t\simeq 1/g$ is much less than the time $1/\Delta{}E$ specified by the uncertainty relation as the upper limit.

During the process of energy transfer to the nucleus and back, the electron shell ``breathes'', periodically decreasing and increasing in size. The effect can be quite significant. For example, in low-charge thorium ions, the size of the 7$s_{1/2}$ shell is (3--4)$a_B$, and associated with it by the M1 transition the 8$s_{1/2}$ shell has sizes of (7--8)$a_B$. A similar situation will be for other shells. Thorium-229 ions trapped in a Paul trap will scatter significantly more light in the excited state than in the ground state. The flickering of ions at a frequency of $10^8$--$10^9$ Hz is quite reliably recorded by modern electronics.

Let us turn to another interesting aspect of our problem. The width of the isomeric nuclear state $\Gamma_N$ corresponds to a half-life of tens of minutes. Therefore, the system ``electron shell of the Thorium ion with a doublet of the ground state levels of the $^{229}$Th nucleus'' can be considered as a nuclear quantum battery, where the electron shell plays the role of a donor, and the nucleus --- an acceptor. This approach allows us to look at the excitation of the low-lying isomer $^{229m}$Th(3/2$^+$, 8.4~eV) as charging a quantum battery.

A large number of publications are devoted to quantum batteries (see, for example, \cite{Ficek-02, Gonzalez-Tudela-11, Ren-17, Farina-19, Cortes-22, Downing-23, Ahmadi-24} and references therein). Here we consider the ``charging'' of $^{229m}$Th by coherent excitation the thorium ion electron shell by resonant laser radiation followed by coherent energy transfer to the nucleus. This is a coherent analogue of the well-known excitation process through an electron bridge, which was first proposed in the works \cite{Tkalya-92-JETPL, Tkalya-92-SJNP} and subsequently considered in a large number of publications \cite{Kalman-94, Porsev-10-PRL,Karpeshin-17,Bilous-18-NJP, Borisyuk-19-PRC, Bilous-20, Dzyublik-20, Dzuba-25-PRA-L,Dzuba-25-PRA,Shigekawa-26,Wense-26}.

Let us add to the Hamiltonian (\ref{eq:Hamiltonian}) a term corresponding to the coherent interaction of the system with laser radiation
\begin{equation*}
\sum_{i=A,N} \Omega_i \left(S^+_i e^{-i\omega_Lt} + S^-_i e^{i\omega_Lt}\right)\,,
\label{eq:Hamiltonian_laser}
\end{equation*}
where $\Omega_i$ is the Rabi frequency, $\omega_L$ is the energy of photons of laser radiation tuned to resonance with the atomic transition $\omega_L = \omega_A$ (and in our case with the nuclear transition too $\omega_A = \omega_N = E_{\text{is}}$).

Due to the smallness of the nuclear magnetic moment of the transition compared to the atomic moment $\mu_{\text{is}\rightarrow\text{gr}}\ll \mu_A$, the condition $\Omega_N \ll \Omega_A$ is satisfied. This allows us to neglect the coherent interaction of laser radiation with the nucleus. Additionally let the decay of the atomic state $\Gamma_A$ be negligible. Then the probability of detecting the system in the state $A_eN_e$ under the initial condition $\rho_{A_gN_g}(t=0) =1$ is determined by the formula
\begin{eqnarray}
\rho_{A_eN_e}(t) = \left(1-\frac{\Omega_A^2}{\left(\frac{g}{2}\right)^2} \sin^2\left(\frac{g}{2} t\right)\right)
\sin^2\left(\frac{\Omega_A}{g}\Omega_A t\right).
\label{eq:PNeAe}
\end{eqnarray}

\begin{figure}
 \includegraphics[angle=0,width=0.95\hsize,keepaspectratio]{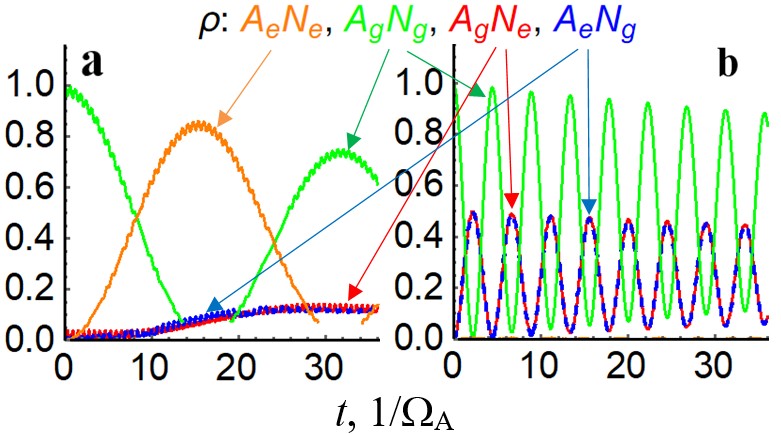}
 \caption{a) The probability to find the system in different states for the ratio of constants $g=10\Omega_A$, $\Gamma_A = \Omega_A/50$ and $\omega_L=E_A$ under the initial condition $A_gN_g(0)=1$ (see explanations in the text). b) The probability to find the system in the same states, but with $\omega_L=E_A \pm g$.}
 \label{fig:Fig4}
\end{figure}

The atomic and nuclear systems oscillate between the ground $A_gN_g$ and excited $A_eN_e$ states with a circular frequency $\Omega_{\text{eff}}=\Omega_A^2/g$. Note that the nucleus and the electron shell are excited primarily synchronously $A_gN_g\leftrightarrow{}A_eN_e$ with a ``low'' frequency $\Omega_{\text{eff}}$ and high excitation probability. The second term in Eq.~(\ref{eq:PNeAe}) describes the high-frequency ripples on the graphs in Fig.~\ref{fig:Fig4}a.

On the other hand the energy of the laser photons $\omega_L = E_A$ is no longer resonant due to the shift in the energy of the states. The new resonant frequencies of the laser radiation are $\omega_L = E_A \pm g$. Tuning to these frequencies leads to the situation shown in Fig.\ref{fig:Fig4}b. Here, excitation of all states occurs at a frequency close to $\Omega_A$, and the maximum probability of the nucleus excitation does not exceed 50\%.

Of course, a laser radiation and a chemical environment influence both the positions of the energy levels (especially electron levels) and the transition energy. But given the existing uncertainty in the energies of the states, this shift can work in both directions, namely, both toward and away from resonance. That is, it can have both a positive and a negative effect, but it is impossible to predict at the moment which one.

Note that in $^{229}$Th$^+$ the excitation process can quickly cease as a result of additional ionization due to the transfer of the total energy 2$E_{\text{is}}$ of the synchronously oscillating nucleus and electron shell to one of the valence electrons.

%
%

For numerical estimates in the $^{229}$Th$^{39+}$ ion, we use Torrey`s solution of the optical Bloch Equations for laser radiation interacting with the electron shell (see in \cite{Wense-20}). Let the laser have a continuous radiation power $P_L$ with a line width $\Delta{}\omega_L \simeq 10 \Gamma_A$. The radiation is focused into a spot with a diameter of $10^{-2}$~cm. Oscillations with effective frequency $\Omega_{\text{eff}} \simeq 3\times10^{-15}$~eV and period $T_{A_gN_g\leftrightarrow{}A_eN_e}\simeq 1$~s begin at $P_L\simeq10^{-5}$~W, when $\Omega_A \simeq 10^{-10}$~eV. Besides, there is no need for the nonreciprocity condition \cite{Ahmadi-24}, which ensures a directed energy flow from the atomic shell to the nucleus. In the scheme presented, it is sufficient to stop the irradiation process at the desired moment.

To summarize. Among the huge number of transitions in the electron shells of ions $^{229}$Th$^{n+}$, there may exist magnetic dipole transitions with an energy close to the energy of the nuclear isomeric transition $3/2^+(E_{\text{is}})\rightarrow 5/2^+(0.0)$. In such an atomic-nuclear system, in the case of relatively large interaction energy of the electron and nuclear currents $E_{\text{int}}$ compared to the radiative width of the electron transition $\Gamma_A$, a multiple coherent transfer of excitation energy from the electron shell to the nucleus and vice versa becomes possible even for significant energy mismatches $\Delta{}E$ between the electron and nuclear transitions $\Gamma_A \lesssim \Delta{}E \lesssim \Re(E_{\text{int}})$. The frequency of these oscillations is $10^8-10^9$~Hz, and the duration can reach $10^{-5}-10^{-4}$~s. Such ``breathing'' of the electron shell can be observed in an ion trap by the intensity of the scattered light. Furthermore, accelerated decay of $^{229m}$Th by orders of magnitude occurs due to the effect of the electron shell spontaneous emission on the emission rate of the nucleus. On the other hand, this atomic-nuclear system is two qubits with strong coupling, spatially nested one inside the other. When coherent laser radiation is applied to an atomic subsystem, synchronous coherent oscillations of the population of the atomic and nuclear excited states begin. One can consider this object as the example of a nuclear quantum battery, for the charging of which (that is, the excitation of a nuclear isomeric state) one can use known methods of charging quantum batteries.

The author thanks the staff of the Laboratory of Optics of Complex Quantum Systems of the P.N. Lebedev Physical Institute of the Russian Academy of Sciences, and especially Prof. A.A. Golovizin and Dr. N.O. Zhadnov for a number of important advices that allowed to understand better results of the work. The work was carried out with the support of the Russian Science Foundation grant No. 24-12-00053 and within the framework of the scientific program of the National Center for Physics and Mathematics, section 6 ``Nuclear and Radiation Physics''.

%


\end{document}